# Quantum technologies with hybrid systems

G. Kurizki[1], P. Bertet[2], Y. Kubo[2], K. Mølmer[3], D. Petrosyan[4,5], P. Rabl[6], J. Schmiedmayer[6]

[1] Department of Chemical Physics, Weizmann Institute of Science, Rehovot 76100, Israel
[2] Quantronics Group, SPEC, CEA-Saclay, 91191 Gif-sur-Yvette, France
[3] Department of Physics and Astronomy, Aarhus University, DK-8000 Aarhus C, Denmark
[4] Aarhus Institute of Advanced Studies, Aarhus University, DK-8000 Aarhus C, Denmark
[5] Institute of Electronic Structure and Laser, FORTH, GR-71110 Heraklion, Crete, Greece
[6] Vienna Center for Quantum Science and Technology, Atominstitut, TU-Wien, A-1020 Vienna, Austria

**An extensively pursued current direction of research in physics aims at the development of practical technologies that exploit the effects of quantum mechanics. As part of this ongoing effort, devices for information processing, secure communication and high-precision sensing are being implemented with diverse systems, ranging from photons, atoms and spins to mesoscopic superconducting and nanomechanical structures. Their physical properties make some of these systems better suited than others for specific tasks; thus, photons are well suited for transmitting quantum information, weakly interacting spins can serve as long-lived quantum memories, and superconducting elements can rapidly process information encoded in their quantum states. A central goal of the envisaged quantum technologies is to develop devices that can simultaneously perform several of these tasks, namely, reliably store, process, and transmit quantum information.** *Hybrid quantum systems* **composed of different physical components with complementary functionalities may provide precisely such multi-tasking capabilities. This article reviews some of the driving theoretical ideas and first experimental realizations of hybrid quantum systems and the opportunities and the challenges they present and offers a glance at the near- and long-term perspectives of this fascinating and rapidly expanding field.**

During the past several decades, quantum physics has evolved from being primarily the conceptual framework for the description of microscopic phenomena to providing inspiration for new technological applications. A range of ideas for quantum information processing [1] and secure communication [2,3], quantum enhanced sensing [4-8] and the simulation of complex dynamics [9-14] has given rise to expectations that society may before long benefit from such quantum technologies. These developments are driven by our rapidly evolving abilities to experimentally manipulate and control quantum dynamics in diverse systems, ranging from single photons [2,13], atoms and ions [11,12], and individual electron and nuclear spins [15-17] to mesoscopic superconducting [14,18] and nanomechanical devices [19,20]. As a rule, each of these systems can execute one or a few specific tasks, but no single system can be universally applied in all envisioned applications. Thus, photons are best suited for transmitting quantum information, weakly interacting spins may serve as long-lived quantum memories, and the dynamics of electronic states of atoms or electric charges in semiconductors and superconducting elements may realize rapid processing of information encoded in their quantum states. The implementation of devices that can simultaneously perform several or all of these tasks, e.g., reliably store, process, and transmit quantum states, calls for a new paradigm: that of hybrid quantum systems (HQSs) [15,21-24]. HQSs attain their multi-tasking capabilities by combining different physical components with complementary functionalities.

Many of the early ideas for HQSs emerged from the field of quantum information processing and communication (QIPC) and were, to a large extent, inspired by the development of QIPC architectures in which superconducting qubits are coupled to high-quality microwave resonators [18,25]. Superconducting qubits are very-well controlled quantum systems [26,27], but in contrast to atoms, they suffer from comparatively short coherence times and do not couple coherently to optical photons. A microwave resonator, such as, for example, a lumped-element LC-circuit or coplanar waveguide (CPW) resonator, can serve as an interface between superconducting qubits and also between superconducting qubits and other quantum systems with longer coherence times and optical transitions [18,22,23,28]. It has thus been proposed to couple superconducting qubits, via a "microwave quantum bus", to ions [29], atoms [30-32], polar molecules [33], electrons confined above a liquid helium surface [34], and spin-doped crystals [15,35-37]. With the recent advances in the control of micro- and nanomechanical systems [19,20], the use of a "mechanical quantum bus" has also been identified as an alternative promising route for interfacing and communicating between various quantum systems [24,38]. Here one exploits the ability of functionalized resonators to respond sensitively to weak electric, magnetic or optical signals while still being sufficiently well isolated from the environment to enable coherent transmission of quantum states.

Nearly a decade after the initial proposals, the HQS idea has developed into a rapidly growing interdisciplinary field of research. The basic working principles of HQS have by now been demonstrated in several experiments, and a growing number of theoretical and experimental research activities is presently devoted to further exploration of this approach. It is thus timely to assess the current status of the field and highlight some of its most promising short- and long-term perspectives. In what follows, this is done by reviewing the overarching goals and focusing on a few examples of HQS representing some of the most active research directions in this field. References are provided where the reader may find further information on this field.

## General Concepts and Experimental Implementations of HQSs

A necessary prerequisite for realizing a functional HQS is the ability to communicate quantum states and properties between its different components with high fidelity. For two physical systems A and B, this requires an interaction Hamiltonian $H_{AB}$ which either conditions the evolution of one system on the state of the other, or drives, in a correlated fashion, transitions in the two systems. In most of the examples discussed below, we will encounter interaction Hamiltonians of the form

$$H_{AB} \simeq \hbar g_{eff}(a^+ b + b^+ a), \qquad (1)$$

where $a$ ($a^+$) and $b$ ($b^+$) are de-excitation (excitation) operators, which, in a generic sense, cause transitions among states within systems A and B, respectively. The product $a^+ b$ thus indicates a swap process in which an excitation in one system is accompanied by a de-excitation in the other. With appropriate identification of the operators, Eq. (1) represents interacting systems such as quantum optical fields coupled by a beam splitter, atoms interacting resonantly with a cavity field, and naturally occurring spins interacting via their magnetic moments.

If systems A and B have very different physical properties, it may be difficult to identify appropriate degrees of freedom which experience interactions of the form of Eq. (1). One obstacle may arise from the effective coupling strength $g_{eff}$ being weak due to inadequate spatial (or impedance) matching between the systems. The couplings between microscopic systems, such as the spin-orbit and spin-spin interactions responsible for fine and hyperfine structure in atoms and molecules, are relatively strong because the electrons and nuclei are confined within Ångström distances. Alternatively, mesoscopic superconducting qubits may strongly couple to each other because of the large electric dipoles associated with the spatial extent of the region traversed by the sustained electric currents. But when a single atom, ion or electron is placed near a micrometer- or millimeter-sized superconducting system, the coupling between the two is several orders of magnitude weaker. Another common challenge is rooted in the difference of the energy scales in the systems that we intend to couple. Even in the presence of a strong interaction, the swap process described by $H_{AB}$ will not take place if it does not conserve energy, meaning that the excitation energies of the two sub-systems are very dissimilar. Much of the research in the field of HQS is devoted to overcoming these obstacles.

In Fig. 1 we illustrate various candidate components of HQSs, characterized by their typical Bohr excitation frequencies (vertical axis) and their coherence times (horizontal axis). The location of each system on the horizontal axis of the figure identifies the tasks that are best delegated to that component of a HQS: e.g., spins are useful for storage while superconducting qubits may be more practical for rapid processing of quantum states. The coherence time $T_2$, i.e. the time over which quantum superposition states survive, determines the minimal coupling strength required for a HQS component to function with sufficiently high fidelity: The (effective) coupling rate $g_{eff}$ between a pair of systems A and B must be large enough to allow quantum state transfer between them within the shortest coherence time of the two, $g_{eff} T_{2\,min} \gg 1$.

The arrows connecting different components of HQS in Fig. 1 are labeled by approximate values of $g_{eff}$ that can be realistically achieved with present-day technology. Some of the larger coupling strengths in Fig. 1 seem to contradict our observation concerning the weak coupling between very different physical systems. This contradiction is resolved by noting that we consider the coupling of the mesoscopic system via light or microwave fields to ensembles rather than to single atoms or spin dopants (see below). The red and blue arrows in the figure indicate the single-system and ensemble coupling strengths, respectively. Figure 1 also shows various examples for the coupling of systems with strongly dissimilar excitation energies (dashed lines). In such cases, the coupling mechanism involves an external source or sink, such as a (classical) laser or microwave field, which bridges the energy mismatch to make the processes described by Eq. (1) resonant when they are accompanied by absorption or stimulated emission of photons. This coupling mechanism applies to the well-known processes of laser-assisted optical Raman transitions in atoms and molecules. In optomechanics [20], parametric coupling via an applied control field is used to bridge the energy difference between mechanical vibrational modes and optical photons and to enhance the interaction strength $g_{eff}$.

We next present in more detail the specifics of different systems sketched in Fig. 1, and describe some of the ideas for their hybridization.

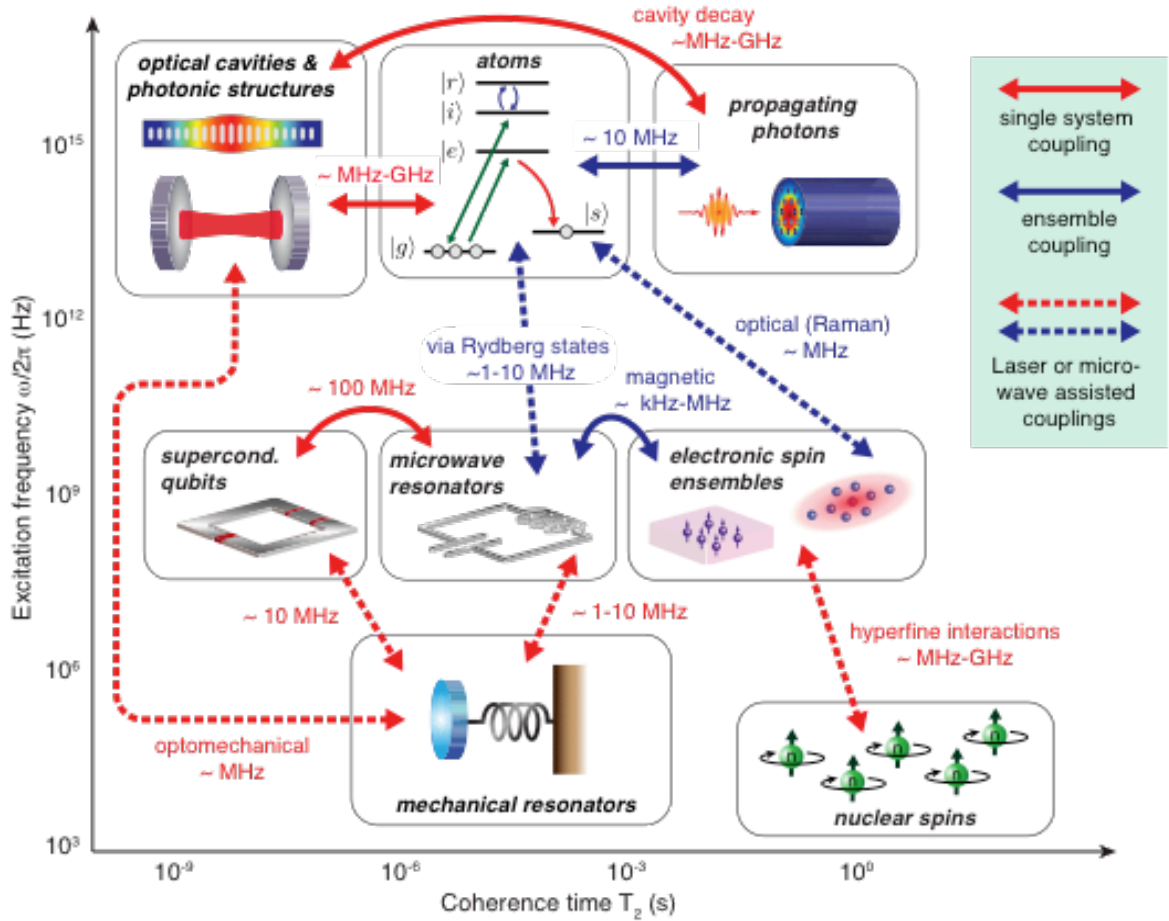

**Fig. 1.** HQS overview. The diagram shows a selection of physical systems that represent components of HQS with different functionalities. The individual systems are positioned in the diagram according to their characteristic excitation frequencies (vertical axis) and coherence times (horizontal axis). The arrows indicate possible coupling mechanisms and the corresponding coupling strength $g_{eff}$ that can be realistically achieved. The red and the blue arrows represent the coupling between single systems and the coupling to and between ensembles, respectively. Couplings represented by dashed lines are assisted by additional classical laser or microwave fields to gap the apparent mismatch of the excitation energies. See text for more details.

**Spin-ensemble quantum memories for superconducting qubits.**

The simplest superconducting qubit is an electrical LC circuit (resonator) in which the inductance is replaced by a non-linear Josephson junction. The excitation spectrum of this resonator then becomes anharmonic at the single-quantum level and can thus behave as an effective two-level system [26,27]. Due to their macroscopic size, typically between 100μm and 1mm, superconducting qubits possess a large electric dipole moment, and hence they couple strongly to the microwave field of a CPW resonator [18]. This strong coupling allows efficient qubit readout, as well as rapid (100 ns in duration) exchange of photons with the resonator which can mediate high-fidelity quantum gates between different qubits [25]. The strong interactions of superconducting qubits with the environment, however, lead to modest coherence times, of the order of 10-100 μs at best [39-40].

On the other hand, due to the small magnetic moments, electronic or nuclear spins interact only weakly with their environment, and even at room temperature spins embedded in a solid-state matrix can exhibit coherence time reaching seconds for electron spins [41] and many minutes, up to hours, for nuclear spins [42]. At the same time, spins can be densely packed since their mutual influence is weak. Thus, spin ensembles provide a natural storage medium for quantum states.

The prospect of combining fast superconducting processing qubits with long-lived spin quantum memories has led to some of the first HQS proposals [35,36], in which spins and qubits are coupled to the same CPW microwave cavity and storage of a qubit state is mediated by a cavity photon which is absorbed by the spins. The magnetic coupling strength between a single spin and a microwave photon is only about $g \sim 10$ Hz in typical CPW resonators, which is too small for a photon to be absorbed before it leaks out of the cavity. It has therefore been proposed to employ a large ensemble of $N$ spins which couple to a photon $\sqrt{N}$ times stronger than a single spin. Indeed, the resonant coupling of an ensemble of $N$ spins to a single cavity mode is described by the Tavis-Cummings model with the Hamiltonian

$$H = \hbar g \sum_{n=1}^{N}(c \mid \uparrow>_n<\downarrow \mid + c^+\mid \downarrow>_n<\uparrow \mid) = \hbar g \sqrt{N}(c\, S^+ + c^+ S\,), \qquad (2)$$

where $c$ ($c^+$) denotes the annihilation (creation) operator of the cavity field, and the operator $S^+ = \frac{1}{\sqrt{N}}\sum \mid \uparrow>_n<\downarrow \mid$ creates a collective spin excitation distributed over the entire ensemble. Equation (2) clearly reveals the collective enhancement factor $g_{\text{eff}} = g\sqrt{N}$ of the effective coupling rate between the cavity mode and the spin ensemble which, for $N=10^{11}$-$10^{12}$, can exceed both the spin decoherence rate $1/T_2$ and typical cavity damping rates $\kappa \approx 10^5$ Hz. Following these early proposals, the strong collective coupling regime has been observed in a number of systems, P1 centers in diamond [43,44], nitrogen-vacancy (NV) centers in diamond [45,46] and erbium ions in YSiO2 [47]. Much larger collective coupling constants can be reached if the ensemble of non-interacting spins is replaced by exchange-coupled electron spins as provided by insulating ferromagnets such as YIG, providing a density of electron spins up to 4 orders of magnitude larger [48-51]. Although the shorter lifetime of the magnonic elementary excitations in these systems (1 ms at best) makes them unsuitable for quantum memory applications, these novel hybrid systems open new possibilities in quantum magnonics and quantum transducers.

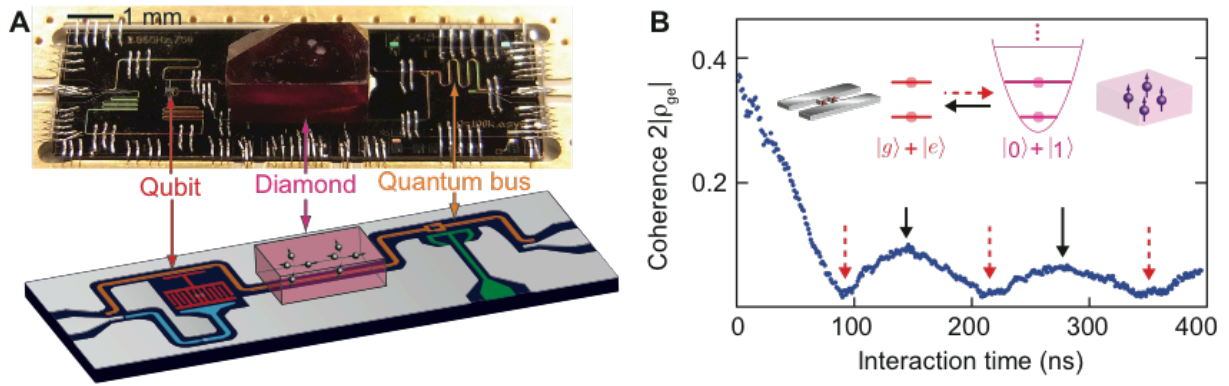

**Fig. 2.** Spin ensemble quantum memory. (A) Photo (Upper) and schematic drawing (Lower) of the hybrid quantum circuit realised in Ref. [53]. A transmon qubit (red) is coupled to an ensemble of NV- center electron spins (pink) via a frequency-tunable quantum bus resonator (orange). (B) Swap oscillations of a coherent superposition of quantum states, (|g> +|e>)/√2 initially prepared in the qubit, showing cycles of storage (dashed arrows) into and retrieval (solid arrows) from the spin ensemble.

Recent experiments have demonstrated the possibility of quantum-state transfer from a superconducting qubit to a spin ensemble [52-55]. The experiment in Ref. [53] is schematically depicted in Fig. 2A: A superconducting qubit is electrically coupled to a SQUID-based CPW resonator, which in turn is magnetically coupled to an NV-center ensemble. An arbitrary superposition $\alpha|g\rangle + \beta|e\rangle$ of the qubit ground ($|g\rangle$) and excited ($|e\rangle$) states is first transferred to the corresponding superposition of the CPW "bus" (resonator) energy states $\alpha|0\rangle + \beta|1\rangle$, with

$|0\rangle$ and $|1\rangle$ denoting the 0- and 1-photon Fock states. This state then mapped onto a superposition $\alpha|\downarrow\rangle + \beta S^+|\downarrow\rangle$ of the spin-ensemble ground state $|\downarrow\rangle$, corresponding to all spins in the ground state, and the collective symmetric single-excitation state $S^+|\downarrow\rangle = (|\uparrow_1\downarrow_2..\downarrow_N\rangle + |\downarrow_1\uparrow_2..\downarrow_N\rangle + \cdots + |\downarrow_1\downarrow_2..\uparrow_N\rangle)/\sqrt{N}$. Typical results are shown in Fig. 2B for the initial qubit state $|\psi\rangle = (|g\rangle + |e\rangle)/\sqrt{2}$ undergoing periodic storage and retrieval cycles in this proof-of-principle demonstration of spin-ensemble quantum memory.

The retrieval signal in Fig. 2B is seen to decay in a few hundred nanoseconds, which is many orders of magnitude faster than coherence times reported for individual NV centers in diamond. This rapid decay stems from the inhomogeneous broadening of the spin ensemble, namely, each spin having a slightly different resonance frequency $\omega_i$ in its specific local environment (formed by other electronic spin impurities, nuclear spins, local strain, etc.). The spread in frequency of the spin ensemble causes the collective single-excitation state $S^+|\downarrow\rangle$ to evolve during time $\tau$ into the state

$$|\Psi(\tau)\rangle = (|\uparrow_1\downarrow_2..\downarrow_N\rangle e^{i\phi_1} + |\downarrow_1\uparrow_2..\downarrow_N\rangle e^{i\phi_2} + \cdots + |\downarrow_1\downarrow_2..\uparrow_N\rangle e^{i\phi_N})/\sqrt{N} \qquad (3)$$

with $\phi_i = -\omega_i\tau$. The accumulation of different phases implies that the terms in the superposition of Eq. (3) no longer interfere constructively and the stored state cannot be retrieved as a cavity photon after storage times exceeding $T_2^* \approx 1/\Delta\omega$, which is determined by the (inhomogeneous) linewidth $\Delta\omega$ of the spin ensemble [53].

The extension of storage time of the solid-state spin-ensemble quantum memories beyond the inhomogeneous linewidth limit is a subject of ongoing research. One remedy to the problem can be to employ stronger couplings, $g_{\text{eff}} \gg \Delta\omega$, to shift the frequency of the hybridized spin wave - cavity excitation out of the continuum of spin transition frequencies. This approach has indeed been demonstrated to lead to a significantly reduced decay of collective excitations in such systems [56]. Another potentially viable approach involves a prior optimal spectral filtering of the spin ensemble [57], which leaves an appropriate subset of the spins to serve as a high-fidelity memory. Finally, because individual spins in the inhomogeneously broadened ensemble evolve unitarily, each with its own transition frequency, the precession direction of all the spins can be reversed by applying well-known refocusing techniques, such as Hahn echo sequences used in magnetic resonance experiments. This refocusing can enable the retrieval of the stored quantum state as a spin-echo long after the ensemble state would have lost its phase coherence due to inhomogeneous broadening [58-61].

On the other hand, inhomogeneous broadening and its effective refocusing by π pulses offer the possibility of using the spin ensemble for multi-mode storage. Because the photonic excitation stored in the ensemble at time $t$ is dephased during time $T_2^*$, another photon can be transferred to a collective spin-excitation mode at time $t+\tau$, if $\tau > T_2^*$. By exploiting the vanishing overlap between any two such storage modes, this transfer process can be repeated several times [58-61]. The number of photon pulses stored within the homogeneous lifetime $T_2$ ($\gg T_2^*$) of individual spins is limited by the time-bandwidth product $T_2/T_2^*$, allowing the storage of up to a few hundred photonic qubits in an ensemble of NV centers [60]. Encouraging experimental results have already been obtained for the sequential storage of several weak microwave pulses and their retrieval tens of microseconds later [61]. Multimode storage has also been demonstrated in a phosphorus-doped silicon crystal [58]. In this experiment, the strong hyperfine interaction between each individual electron and its parent ion was further exploited to convert a sequence of collective electron-spin excitations into nuclear excitations, which were then stored and subsequently transferred back to the electron spins and the microwave field after seconds of storage time.

These initial results and ideas indicate the feasibility of a practical spin-ensemble quantum memory, capable of simultaneously storing the states of hundreds of superconducting qubits, for many seconds, and, potentially, for hours. Such memories would constitute a prime example of the benefits of the HQS approach.

**Atomic ensembles as memories and optical interfaces.**
In addition to information processing and storage, quantum states may be used for secure transmission of data. Distribution of data within larger QIPC architectures and practical quantum cryptography can only be achieved using optical photons propagating in free space or optical fibers [2,3]. The implementation of coherent interfaces between quantum memories, processing qubits and "flying" optical qubits is thus of general importance in quantum information science.

Isolated atoms have optical (Raman) transitions with excellent properties for coherent absorption and emission as well as storage of photons. Therefore the coupling of ensembles of cold trapped atoms [31,32] and molecules [33] to CPW resonators were among the initial proposals for HQSs capable of providing both long storage times and efficient optical interfaces. Compared with solid-state spin ensembles, cooling and trapping atoms in the vicinity of superconducting CPW, however, introduce severe experimental complications, and the realistic number of trapped particles of about $N=10^5$-$10^6$ is much smaller than what is achievable with spin-doped crystals and hence the coupling to atomic spin ensembles is generally weaker [31].

To compensate for the reduced coupling, strong electric-dipole transitions between rotational states in polar molecules [33] or between highly excited Rydberg states of atoms [32,62] can be used. In particular, the transitions in the range of tens of GHz between circular Rydberg states with huge dipole moments have been employed [63] to strongly couple Rydberg atoms to single microwave photons in 3D resonators. Preliminary findings show that similar coupling can also be realized in on-chip CPW cavities [64,65]. Reference [32] details a scheme where a cloud of atoms initially prepared in the ground state $|g>$ is coupled to a collective Rydberg excitation state via a two-photon process involving an optical pump field and a single photon of a CPW resonator. Overall, this scheme realizes an interaction Hamiltonian of the generic form given by Eq. (1),

$$H = \hbar g\sqrt{N}(c\, R^+ + c^+ R), \qquad (4)$$

where the operator $R^+ = \frac{1}{\sqrt{N}} \sum |r>_n <g|$ creates a symmetric (collective) excitation of the Rydberg state $|r>$ and $g = \Omega_{gi}\, \eta_{ir}/\delta$ is the effective two photon coupling rate. This rate is proportional to the Rabi frequency $\Omega_{gi}$ of the optical pump field between the ground state $|g>$ and an intermediate Rydberg state $|i>$ and the Rabi frequency $\eta_{ir}$ of a single microwave photon interacting with the Rydberg states $|i>$ and $|r>$, and is inversely proportional to the detuning $\delta \gg \Omega_{gi}, \eta_{ir}$ from the intermediate state. With strong enough pump field $\Omega_{gi}$ and the very large dipole matrix element ($\eta_{ir} \sim n^2$), such as obtained for the transition between neighbouring Rydberg states with high principal quantum number $n \cong 70$, the necessary effective coupling strength of $g_{\text{eff}} = g\sqrt{N} \gtrsim 1$ MHz can already be reached with a reasonable number of $N=10^6$ atoms.

Once the state transfer between the microwave photon and the atoms is complete, additional optical transitions can be used to transfer the collective Rydberg excitation to a long-lived spin excitation in the ground-state hyperfine manifold and back, or to map it onto a propagating photon mode. In free space, stimulated Raman techniques, such as electromagnetically induced transparency [66] enable a coherent and reversible conversion of collective spin excitations into photons in well-defined spatiotemporal modes. This process requires, however, large optical depth of the medium, *OD = σρl > 1*, where $\sigma$ is the single-atom absorption cross-section, $\rho$ is the atom density and $l$ is the length of the medium [67]. By placing the atomic medium in an optical resonator with high finesse $F$, the transfer efficiency can be further increased, reaching the optimal value of $C/(1 + C)$ [68]. Here $C = F \times OD$ is the optical cooperativity which represents the key figure of merit for the optical interface. In view of low atomic densities, the integration of atomic traps with on-chip photonic structures [69,70] is a promising experimental approach toward achieving hybrid interfaces with $C \gg 1$.

Trapping and cooling of atoms in chip-based traps is a well-established technique, which is, however, technically demanding at cryogenic temperatures and close to a superconducting surface. Nevertheless, magnetic trapping and even cooling of a cloud of atoms to quantum degeneracy (BEC) above a superconducting chip has been demonstrated [71,72]. New ways for using superconducting resonators and circuits directly for trapping are currently being explored [73]. Other experimental approaches for atomic HQSs are being pursued [74], where atoms are optically trapped in the evanescent field of a tapered fibre [75]. This proximity of the atoms to the fibre would enable high-fidelity conversion of atomic quantum states into propagating photons in the fibre while minimizing the perturbation of the superconducting circuit because of the localization of the trapping and Raman laser beams.

In parallel to these experimental efforts on atomic microwave-to-optics interfaces, other approaches employing optical and microwave transitions in spin-doped crystals are being developed [76,77]. The goal here is to achieve a maximal overlap between the spin excitations created by the respective optical and microwave modes while minimizing the absorption of optical photons on the superconductor. Using for this purpose larger 3D microwave resonators, instead of planar CPW, may be beneficial, because both the optical and microwave modes can have their maxima at the center of such resonators, away from the superconducting walls [77]. The reduced magnetic or optical coupling to a single spin is compensated in this approach by the larger number of particles that can be enclosed within the microwave mode volume [78-80].

**Mechanical quantum transducers.**
As an alternative to optical and microwave photons, quantum information can be converted and transmitted via quantized mechanical vibrations of opto- and nanomechanical systems [19,20]. High-Q micro- and nano-mechanical resonators, such as tiny cantilevers or suspended membranes, respond very sensitively to applied forces, which makes them suitable for diverse applications relying on the detection of weak signals. When applied at the level of single quanta, the same principle of force sensing - for example, the conversion of a weak magnetic force into a detectable electric or optical signal [81,82] - opens up new possibilities for mechanically interfacing quantum systems of different types [24]. The recently demonstrated cooling of mechanical vibrational modes close to the quantum ground state [83-85] and the current experimental efforts to couple mechanical resonators to superconducting circuits [83,86,87], atoms [88] and spins [89,90] are initial important steps in this direction.

The basic ideas and potential applications of mechanical quantum transducers are illustrated in Fig. 3. The setup shown in Fig. 3A depicts a mechanical spin transducer, where localized electronic spin qubits are coupled to the quantized motion of a magnetized vibrating tip. In the presence of strong magnetic field gradients, the motion of the tip modulates the Zeeman splitting of the spin eigenstates below the tip and results in a spin-resonator interaction of the form [91]

$$H_{\text{int}} = \hbar\lambda(b + b^+)\sigma_z. \quad (5)$$

Here $b$ is the annihilation operator for the mechanical oscillator mode, $\sigma_z$ is the Pauli operator for the spin and $\lambda$ is the coupling strength per phonon. Unlike Eq. (1), $H_{\text{int}}$ does not describe a resonant exchange of excitations. Instead, it represents a spin-dependent force, which evolves an initial spin superposition state into an equivalent superposition of displaced mechanical states.

If the resonator is electrically charged, the weak magnetic moments of the spins are effectively amplified via this process into large electric dipoles, enabling, for example, strong electric interactions between separated spin qubits. Similar principles underlie the realization of various other mechanical hybrid systems in which spins or superconducting qubits are mechanically coupled with each other or interfaced with photons [92], trapped ions [93,94] or atomic systems [88,95-97].

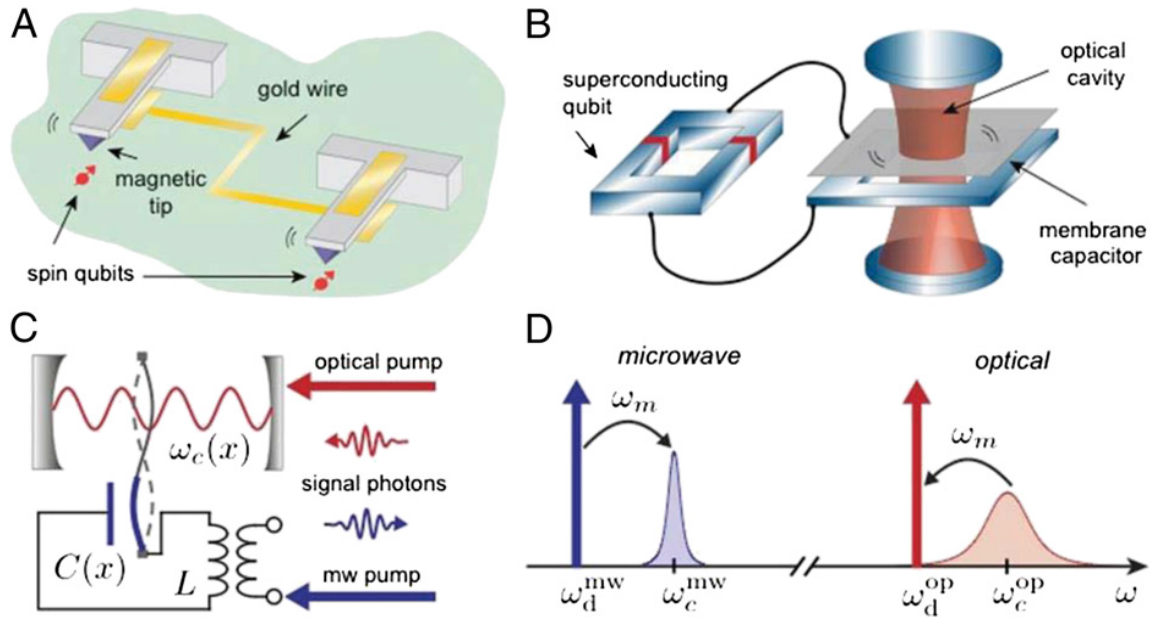

**Fig. 3.** Mechanical quantum transducers. (A) A magnetized mechanical resonator is coupled to a localized electronic spin qubit and converts small spin-induced displacements into electric signals. Thereby spin qubits can be "wired up" electrically or coupled to other charged quantum systems. (B) Illustration of an OM interface between a superconducting qubit and optical (flying) photons. Here the mechanical system, represented by a semi-transparent membrane, simultaneously acts as capacitor *and* an optical reflector. (C) Experimental setup used in Ref. [108] to implement an optomechanical microwave-to-optics interface via simultaneous coupling of a partially metallized membrane to an optical cavity and an LC circuit. (D) A signal photon of frequency $\sim \omega_c^{op}$ enters the optical cavity and is down-converted via the driven, parametric OM interaction into a phonon of frequency $\omega_m$. Then, via an equivalent process, this mechanical excitation is up-converted again into a microwave photon of frequency $\sim \omega_c^{mw}$ in the LC circuit. Via this mechanism and its reverse, quantum information encoded in microwave excitations of a superconducting qubit or LC resonator can be coherently transferred into optical signals for long-distance quantum communication.

First experiments - still in the classical regime - have shown that micromechanical oscillators can be magnetically coupled to hyperfine states of cold atoms [88] or individual impurity spins [89,90]. For a coherent coupling of two or multiple qubits via mechanical channels, it is necessary to reach the regime of strong (mechanical) cooperativity, $C_m = \lambda^2 T_2 T_m > 1$ [91]. Here $T_2$ is the qubit coherence time and $T_m^{-1} = k_B T/(\hbar Q)$ is the characteristic mechanical decoherence rate, where $Q$ is the quality factor of the resonator mode, $T$ is the support temperature and $k_B$ is Boltzmann's constant. Simple estimates show that for spin qubits the condition $C_m > 1$ can be realistically achieved using state-of-the-art mechanical resonators with $Q \sim 10^5 - 10^6$ and working at $T \leq 1$ K temperatures [90,91]. For superconducting qubits the electrostatic interaction with nanomechanical resonators can be significantly stronger [98]. Initial experiments accessing the full strong-coupling regime $\lambda > T_2^{-1}, T_m^{-1}$ have already been performed [83,87].

Nano-mechanical systems are of particular interest for the development of a universal optomechanical (OM) transducer for coherently interfacing optical and microwave photons [38,99-101] (Fig. 3C and D). In OM systems, the frequency $\omega_c$ of an optical cavity mode is modulated by the motion of a mechanical resonator with mechanical frequency $\omega_m$. Common examples of OM systems include Fabry-Perot cavities with a movable end-mirror or with a semi-transparent membrane placed inside the cavity [20]. Nano-photonic systems [102,103] and photonic bandgap structures [99,104] also exhibit similar interactions. The OM system is described by the Hamiltonian

$$H = \hbar\omega_c c^+ c + \hbar\omega_m b^+ b + \hbar g_0 c^+ c\,(b + b^+), \qquad (6)$$

where $c$ is the annihilation operator for the optical mode. The first two terms in this equation represent the unperturbed energies of the optical and mechanical modes, respectively, and the third term describes the radiation pressure coupling. The coupling constant $g_0$ is the optical frequency shift per vibrational quantum, which is typically very small. However, the OM interaction can be parametrically amplified by driving the cavity with a strong laser of frequency $\omega_d$. In this case, the total field inside the cavity is $c = [\alpha(t) + \delta c]\exp(-i\omega_d t)$, where $\delta c$ represents the quantum fluctuations around a large classical field amplitude $\alpha(t)$. By choosing the resonance condition $\omega_c = \omega_d + \omega_m$, the dominant OM coupling term is then analogous to a beam-splitter interaction [20],

$$H_{\mathrm{OM}} \simeq \hbar G(t)(\delta c^+ b + b^+ \delta c), \qquad (7)$$

where, similarly to an anti-Stokes scattering process, low-frequency mechanical excitations are up-converted by the driving field into optical signal photons and vice versa. The effective coupling $G(t) = \alpha(t) g_0$ is then enhanced and controlled by the external driving field.

In the microwave domain, an analogous coupling arises when mechanical oscillations modulate the capacitance of a superconducting CPW resonator or LC circuit [84,105]. Here too, the coupling is amplified and controlled by a strong microwave driving field which bridges the frequency difference between the circuit and mechanical resonances. It is important that due to the underlying parametric interaction, the effective photon-phonon interface in Eq. (5) does not rely on the absolute frequency of the optical or microwave mode, which in both cases is compensated by the frequency of the external driving field (Fig. 3D). Therefore, by using a single mechanical membrane both as a mirror and a capacitor, an effective optics-to-microwave interface is achieved, whereby microwave photons are converted to phonons and successively to (flying) optical photons.

Various designs for the experimental implementation of coherent microwave-to-optics transducers are currently being explored. An efficient OM conversion between optical and microwave signals has already been demonstrated in both room-temperature [106,107] and cryogenic [108] environments. Despite many obstacles that hinder fully coherent operation of such experiments at a single-photon level, the prospects for OM quantum interfaces between solid-state, atomic and optical systems are very promising.

## Outlook: Quo Vadis, Quantum Hybridium?

Hybrid quantum systems are still a long way from implementing general quantum information processing and communication tasks with the fidelities needed for practical applications. The integration of very different physical components presents technological and scientific challenges that are absent when controlling each component individually. However, the experimental realizations of HQSs described above show that these obstacles may be overcome. Protocols with controlled time-dependent couplings are currently being investigated to optimize quantum state transfer speed and fidelity. The shape of the coupling fields can be tailored to the temporal response of the noisy (decohering) environments [109], and quantum measurements and feedback may also be applied [110], in the effort to reduce or eliminate the effect of dissipation at HQS interfaces. There is little doubt that continued progress along these directions will enable hybrid, multi-tasking quantum technologies of increasing sophistication.

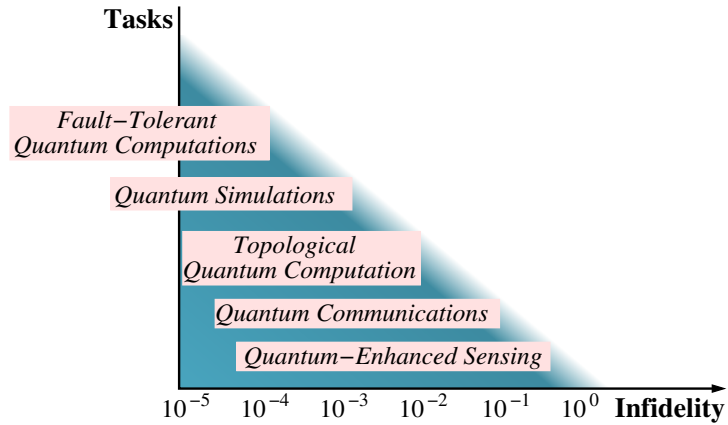

**Fig. 4.** Potential applications that can be performed by a HQS with achievable infidelity 1-*F*.

To provide an overview of tasks for which HQS may be used in the near- and more-distant future, we indicate in Fig. 4 the infidelities tolerated by various potential technological applications of quantum effects. Obviously, the ultimate goal of realizing a large-scale quantum computer including algorithmic quantum error correction is beyond current capabilities. At the other, more modest, end of the scale are communication and sensing which may function, albeit at a lower rate, even if the coupling process has a low fidelity or is heralded with low success probability. Here HQSs are expected to have a great impact already soon. For example, the unprecedented level of control and low-noise amplification at microwave frequencies achieved in superconducting circuits has already been used to detect electron-spin resonance at the level of few-excitations [111]. Similarly, optomechanical transducers are being explored for low-noise optical detection of weak radio-frequency signals [107]. More generally, the anticipated ability of HQSs to transfer highly non-classical (entangled) states between different physical platforms may extend quantum-enhanced sensing schemes to systems where such a high level of quantum control is *a priori* not available. The mapping of squeezed microwave or optical fields onto spin ensembles for magnetometry, or preparation of nanomechanical sensors in highly-sensitive quantum superposition states are possible applications along these lines.

Turning from practical applications to more fundamentally oriented research, the scaling-up of HQSs may offer new possibilities to simulate and study complex phenomena in quantum many-body systems. Especially, the combination of different systems with optimized coherent and tailored (engineered) dissipative properties may be used to investigate issues related to non-equilibrium phase transitions in open quantum many-body systems [10,112-114]. Such systems are currently difficult to realize in the laboratory, but analyses of dissipative Dicke-type lattice models with HQS arrays of spin ensembles and superconducting circuits [115-118] show the potential for achieving scalable systems with sufficiently large interaction strengths.

Overall, the HQS approach concurs with the long-term vision of a "quantum information era", in which quantum information is processed, stored and transmitted in a modular and platform-versatile manner. The fascinating prospects of this research will keep scientists occupied for some time and are likely to stimulate many ideas and motivate experts from research and engineering areas not even mentioned in this review to confront the challenges of HQSs.